# Application Experiences on a GPU-Accelerated Arm-based HPC Testbed

WAEL ELWASIF, WILLIAM GODOY, NICK HAGERTY, J. AUSTIN HARRIS, OSCAR HERNANDEZ, BALINT JOO, PAUL KENT, DAMIEN LEBRUN-GRANDIE, ELIJAH MACCARTHY, VERÓNICA G. MELESSE VERGARA, BRONSON MESSER, ROSS MILLER, and SARP ORAL, Oak Ridge National Laboratory, USA
SERGEI BASTRAKOV, MICHAEL BUSSMANN, ALEXANDER DEBUS, KLAUS STEINIGER, JAN STEPHAN, and RENÉ WIDERA, Helmholtz-Zentrum Dresden-Rossendorf, Germany
SPENCER H. BRYNGELSON, HENRY LE BERRE, ANAND RADHAKRISHNAN, and JEFFREY YOUNG, Georgia Institute of Technology, US
SUNITA CHANDRASEKARAN, University of Delaware, US
FLORINA CIORBA and OSMAN SIMSEK, University of Basel, Switzerland
KATE CLARK, FILIPPO SPIGA, JEFF HAMMOND, and JOHN E. STONE, NVIDIA Corporation, USA
DAVID HARDY, University of Illinois at Urbana-Champaign, USA
SEBASTIAN KELLER and JEAN-GUILLAUME PICCINALI, Swiss National Supercomputing Center, Switzerland
CHRISTIAN TROTT, Sandia National Laboratories, USA

This paper assesses and reports the experience of ten teams working to port, validate, and benchmark several High Performance Computing applications on a novel GPU-accelerated Arm testbed system. The testbed consists of eight NVIDIA Arm HPC Developer Kit systems built by GIGABYTE, each one equipped with a server-class Arm CPU from Ampere Computing and A100 data center GPU from NVIDIA Corp. The systems are connected together using Infiniband high-bandwidth low-latency interconnect. The selected applications and mini-apps are written using several programming languages and use multiple accelerator-based programming models for GPUs such as CUDA, OpenACC, and OpenMP offloading. Working on application porting requires a robust and easy-to-access programming environment, including a variety of compilers and optimized scientific libraries. The goal of this work is to evaluate platform readiness and assess the effort required from developers to deploy well-established scientific workloads on current and future generation Arm-based GPU-accelerated HPC systems. The reported case studies demonstrate that the current level of maturity and diversity of software and tools is already adequate for large-scale production deployments.

Notice: This manuscript has been authored in part by UT-Battelle, LLC under Contract No. DE-AC05-00OR22725 with the U.S. Department of Energy. The United States Government retains and the publisher, by accepting the article for publication, acknowledges that the United States Government retains a non-exclusive, paid-up, irrevocable, world-wide license to publish or reproduce the published form of this manuscript, or allow others to do so, for United States Government purposes. The Department of Energy will provide public access to these results of federally sponsored research in accordance with the DOE Public Access Plan (http://energy.gov/downloads/doe-public-access-plan).

## 1 INTRODUCTION

Deploying new supercomputers requires continuous evaluation of new platforms and understanding of the trade-offs in porting existing applications to different architectures. High Performance Computing (HPC) applications have grown in complexity over the years while new disruptive architectures, including accelerators, have been introduced to increase the computational capabilities with the additional challenge of keeping power requirements within reasonable limits. With many of the HPC technology players building general purpose or specialised accelerators, it increasingly important to have a concrete understanding of the level of human-time investment required to make applications production-ready on any of these accelerated platforms, as well as the expected performance benefits to be gained with such effort.

Since the introduction of Arm Neoverse IP by Arm Ltd, we have witnessed a steady adoption and increasing number of CPU products based on the Arm Instruction Set Architecture (ISA). Noticeable deployments include the 1.529 Petaflops Sandia Astra-2 (first petascale-class Arm-based HPC system in 2018[1]) and the RIKEN R-CCS Fugaku (first exascale-class Arm-based HPC system in 2020[2]). Fugaku, based on Fujitsu's A64FX Arm-based CPU[3] was also the first systems adopting a CPU supporting Arm Scalable Vector Extension (SVE [33]).

For the emerging field of cloud-based HPC, the Arm-based Graviton processor[4] provides a significant portion of computational resources provisioned by Amazon Web Services (AWS). Currently in its 3rd generation (incluuding a dedicated flavour for HPC-workloads called 3E), the Graviton CPU is based on Arm Neoverse V1 core IP and support Arm SVE SIMD instructions. AWS is not the only hyperscaler interested in deploying Arm CPUs, others like Microsoft and Oracle have started to offer Arm-based instances primarily based on Ampere Computing Altra CPU (not HPC-oriented configurations).

---
[1] https://www.sandia.gov/labnews/2018/11/21/astra-2/
[2] https://www.top500.org/system/179807/
[3] https://www.fujitsu.com/global/products/computing/servers/supercomputer/a64fx/
[4] https://aws.amazon.com/ec2/graviton/

At the very early days of the Arm journey into HPC, deployments of hybrid Arm systems were often custom built and in limited scale. The Montblanc [29] project and the UK Catalyst project have paved the way to more robust and accessible systems. We should note that Fugaku and other high-end HPC Arm platforms rely exclusively on the Arm CPU for their computational power, thus providing a homogeneous CPU-only environment. At the same time, hybrid CPU–GPU systems are becoming the dominant choice for high-end, large-scale leadership facilities (above ∼100 PFlops) due to their performance, computational density, and power efficiency. A platform combining a modern Arm-based CPU with an energy-efficient, high-performance GPU appears to be a natural choice for future computing challenges.

The Oak Ridge National Laboratory in collaboration with NVIDIA pioneered the combined use of Arm CPU and NVIDIA GPU in 2019, using Marvell ThunderX2 CPUs and NVIDIA Volta V100 GPUs. More recent efforts like NVIDIA's Arm DevKit[5] and the upcoming NVIDIA Grace Hopper Superchip[6] combine both high-end, commodity server-class CPUs with performant datacenter GPUs like NVIDIA's A100 and H100.

In this fact-pace evolving landscape of accelerators and heterogeneous systems, assessing as early as possible the viability of any technology and its impact software maturity, code portability, and developer productivity remains a must.

To this end, we present an application-focused assessment of a new multi-node accelerator-based platform based on the NVIDIA Arm HPC Developer Kit, with each compute node equipped with an Arm-based Ampere Computing Altra CPU[7], two NVIDIA Ampere A100 PCIe GPUs[8], and two NVIDIA BlueField-2 network cards [9]. This system represents a deployable on-prem system suitable to validate software and ecosystem readiness. This test bed is part of an experimental HPC cluster facility called Wombat, which is discussed in section 2.

This study makes the following contributions: 1) the first thorough collaborative investigation of a modern GPU-accelerated Arm-based system using production applications; 2) the analysis of the readiness of software tools required to compile selected flagship applications with and without GPU support; 3) Preliminary performance results compared to current GPU-accelerated platforms, primarily ORNL's Summit system; and 4) General conclusions on the readiness of the software ecosystem for the Arm-based platforms.

A representative sample of production-ready community-wide HPC applications has been selected for this evaluation. Basic information on these applications is summarised in table 1. Since the primary goal is to assess porting feasibility and gain an initial baseline of current performance, we used the applications *as-is* without explicit tuning efforts beyond adapting compiler and linker flags used or using vendor-provided optimized libraries.

## 2 WOMBAT TESTBED

### 2.1 Background

Wombat is a small cluster which has been equipped since 2018 with various Arm-based platforms from different vendors. The cluster is deployed and managed by The Oak Ridge Leadership Computing Facility (OLCF) and is freely accessible to users and researchers. The purpose of the cluster is to serve as a testbed for Arm-based AArch64 processors and related technologies within a close-to-production environment. The hosted hardware evolves relatively frequently, but it has always been focused on Arm-based CPUs. Users who request access can use the system to port and validate their applications on different Arm CPU architectures, as well as enabling platform engineers to experience end-to-end integration and configuration aspects of a complex Arm-based HPC system.

At the time of writing, the cluster consists of three set of compute nodes:

(1) 4 `HPE Apollo 70` nodes, each equipped with dual-socket Cavium (now Marvell) ThunderX2 CN9980 processors and two NVIDIA V100 GPUs, connected via PCIe Gen 3 bus.
(2) 16 `HPE Apollo 80` nodes, each equipped with a single-socket Fujitsu A64FX processor.
(3) 8 `NVIDIA ARM HPC Developer Kit` nodes, each equipped with a single-socket Ampere Computing Altra Q80–30 CPU and two NVIDIA A100 GPUs - connected via PCIe Gen 4 bus.

These three types of nodes share a common login node based on dual-socket ThunderX2 CPU, and all nodes are connected via either InfiniBand EDR or HDR to the same Infiniband network.

### 2.2 Programming Environment

This section describes the environment and base software setup of Wombat at the time the evaluation took place in April and May 2022. Once selected, we consciously decide not to constantly vary the environment and create a fixed baseline.

Wombat nodes boot their OS from the network, and all nodes are provisioned with the same pre-built compute image based on CentOS 8.1 with kernel 4.18. Job submission and execution are orchestrated using SLURM.

The compilers and interpreters available include NVIDIA HPC SDK (NVHPC) 22.1, Arm Compiler for HPC 22, CUDA 11.5.1, GNU 11.1, LLVM 15.0.0 with OpenMP offload, Python 3.9.0, and Julia 1.7.0. Networking support is provided by OFED 5.4 and UCX 1.11.1 and although most experiments are single node, OpenMPI 4.1.2a1 is installed for multi-node jobs. NSight Compute SDK, Allinea Forge, and Score-P are available as profilers.

We use Spack [14] for additional third party scientific libraries and tools, including HDF5, OpenBLAS, and Score-P. We did not manually modify any compiler optimization flags used by Spack, instead aiming for an unfiltered out-of-the-box experience. Packages that did not have working Spack recipes were installed individually.

Each application team was responsible for building their respective application, installing extra dependencies, and linking the appropriate libraries.

---

[5]https://developer.nvidia.com/arm-hpc-devkit
[6]https://www.nvidia.com/en-us/data-center/grace-hopper-superchip/
[7]https://amperecomputing.com/processors/ampere-altra
[8]https://www.nvidia.com/en-us/data-center/a100/
[9]https://www.nvidia.com/content/dam/en-zz/Solutions/Data-Center/documents/datasheet-nvidia-bluefield-2-dpu.pdf

## 3 EVALUATION METHODOLOGY

By definition, any testbed may lack some features found in fully developed and optimized production systems. This fact should be taken into consideration when analyzing the performance results presented. The common performance score, Time-to-Solution, is not the primary Figure of Merit. That is, our goal is not to perform a deep dive into the performance of any particular computational kernel(s). Rather, we perform a breadth-first study to comprehensively understand the target platform's readiness. This strategy sets the stage for further refinements and possible iteration on system configuration, aiming for robust deployment in a production environment. These refinements might include varying combinations of Arm-based CPUs, GPUs, and intra-node and inter-node connectivity.

We initiated the evaluation process by inviting 26 application teams with major codes that span different application domains and have a major presence on leading edge HPC platforms such as Summit and Piz Daint. Of the 26 invited teams, 13 applications agreed to participate in the evaluation process, and 10 teams eventually carried out the evaluation work. Table 1 summarizes the final list of applications and their key characteristics. The list covers eight different scientific domains and includes codes written in Fortran, C, and C++. The parallel programming models used were MPI, OpenMP/OpenACC, Kokkos, Alpaka, and CUDA. We did not include changes to the application codes in the porting activities.

The evaluation process primarily focuses on application porting and testing, with less emphasis on absolute performance in light of the experimental nature of the testbed. Application teams were responsible for the basic configuration and build management for their respective application with support for installing needed system-wide packages using Spack as needed. The evaluation took place over two months spanning April and May 2022. Application teams were free to choose the particular use cases to be evaluated for usability and performance on the testbed and to compare such performance with other platforms where the respective codes are regularly deployed.

## 4 APPLICATIONS

| App. Name | Science Domain(s) | Language | Parallel Programming Model(s) |
|---|---|---|---|
| ExaStar | Stellar Astrophysics | Fortran | OpenACC, OpenMP offload |
| GPU-I-TASSER | Bioinformatics | C | OpenACC |
| LAMMPS | Molecular Dynamics | C++ | MPI, OpenMP, KOKKOS |
| MFC | Fluid Dynamics | Fortran | MPI, OpenACC |
| MILC | QCD | C/C++ | CUDA |
| NAMD/VMD | Molecular Dynamics | C++ | Charm++, CUDA |
| PIConGPU | Plasma Physics | C++ | Alpaka, CUDA |
| QMCPACK | Chemistry | C++ | OpenMP offload, CUDA |
| SPECHPC 2021 | Variety of applications | C/C++ Fortran | OpenMP offload, OpenMP |
| SPH-EXA2 | Hydrodynamics | C++ | MPI, OpenMP, CUDA |

**Table 1.** Applications evaluated on the Wombat testbed.

### 4.1 ExaStar

*4.1.1 Background.* The toolkit for high-order neutrino-radiation hydrodynamics (*thornado*) [22] is a Fortran code (F2008) written as a stand-alone module that can be incorporated into ExaStar simulations [17] using the Flash-X multi-physics code.thornado is used to compute the neutrino radiation field with a two-moment model for spectral neutrino transport that evolves moments of the neutrino phase-space distribution function representing spectral energy and momentum densities. We use a Discontinuous-Galerkin (DG) phase-space discretization for space and energy combined with Implicit-Explicit (IMEX) time discretization. Compiler directives (e.g., OpenACC or OpenMP) and GPU-optimized linear algebra libraries (e.g., cuBLAS) offload computation to GPUs.

For the ExaStar exascale challenge problem that we plan to run with Flash-X, neutrino transport evolves >90% of the degrees of freedom in core-collapse supernova simulations and is the largest computational bottleneck. This motivates using stand-alone thornado benchmarks as a tool for evaluating node-level performance. Here, we consider two benchmark problems: Streaming Sine Wave and Relaxation. The Streaming Sine Wave benchmark executes only the *Explicit* piece of the IMEX time integration, while the Relaxation benchmark includes only the *Implicit* component. These two benchmarks accurately capture the computational characteristics of neutrino transport in an ExaStar simulation.

*4.1.2 Porting for functionality and correctness experience.* We required minimal modifications to build and run the thornado benchmark problems on the Wombat system. We were able to use existing Makefiles for the NVIDIA HPC SDK compilers on Summit with slight tweaks to the target processor flags. For the CPU, we changed the flags to link LAPACK and BLAS libraries to use OpenBLAS instead of IBM's optimized implementations for the POWER9 architecture. The CPU and GPU programming models use compiler directives that we need not modify. All benchmarks passed built-in correctness tests based on known analytic solutions.

*4.1.3 Performance and comparisons.* As a baseline, we ran both benchmarks on a single node of the Summit computer at the Oak Ridge Leadership Computing Facility (OLCF). Each Summit node has 2 IBM POWER9 CPUs and 6 NVIDIA Volta GPUs, but for comparisons to the NVIDIA ARM HPC Dev Kit, we limit comparisons to a single CPU or single GPU. For the CPU runs with POWER9, we also test different configurations of Simultaneous Multithreading (SMT). The total number of OpenMP threads is set by the product of the number of cores and hardware threads available. To demonstrate the parallel efficiency of our OpenMP implementation, we also report serial execution times for each CPU. On both systems, we use standard -O2 optimizations and -tp for the target CPU. For benchmarks that report using the GPU, all computation is done on the GPU; the CPU thread is only used to launch kernels and manage data transfer. In both cases, the salient Figure of merit is wall-time (lower is better).

**Streaming Sine Wave.** We report the total wall-time to evolve ten timesteps of the Streaming Sine Wave benchmark for each hardware configuration in table 2. The serial CPU comparison shows a speedup factor of 1.3 (2.5) for the Ampere Altra relative to the POWER9 (ThunderX2). This single-core performance gain is also realized for the multi-core comparison, where we find speedup by a factor of 2.2 (2.8) for Altra relative to POWER9 (ThunderX2). However, we find poor strong scaling of Altra (18% parallel efficiency with 80 threads) relative to POWER9 (42% efficiency with 21 threads). We speculate that this is rooted in the introduction of OpenMP overhead stemming from many small loop nests used in the streaming advection operation. This is further supported by the drop in performance on POWER9 for increasing SMT levels. The Altra+A100 results also exhibit a speedup factor of 1.3 (1.9) relative to the POWER9+V100 (TX2+V100) and a factor of 5.2 relative to the Altra multi-core result.

| CPU | GPU | Cores:SMT:Thrds. | Prog. Model | Time (sec) |
|---|---|---:|---|---:|
| Power9 | None | 1:1:1 | OpenMP | 129 |
| ThunderX2 | None | 1:1:1 | OpenMP | 244 |
| Ampere Altra | None | 1:1:1 | OpenMP | 99.0 |
| Power9 | None | 21:1:21 | OpenMP | 14.8 |
| Power9 | None | 21:2:42 | OpenMP | 17.0 |
| Power9 | None | 21:4:84 | OpenMP | 21.3 |
| ThunderX2 | None | 28:1:28 | OpenMP | 18.6 |
| ThunderX2 | None | 28:2:56 | OpenMP | 17.8 |
| ThunderX2 | None | 28:4:112 | OpenMP | 18.5 |
| Ampere Altra | None | 80:1:80 | OpenMP | 6.72 |
| Power9 | V100 | 1:1:1 | OpenACC | 3.75 |
| ThunderX2 | V100 | 1:1:1 | OpenACC | 5.54 |
| Ampere Altra | A100 | 1:1:1 | OpenACC | 2.96 |

**Table 2.** Comparison of thornado wall-clock times on each platform for the Streaming Sine Wave test problem. All runs used the nvfortran compiler. Green rows indicate NVIDIA ARM HPC Development Kit hardware.

In fig. 1, we break down the relative cost of the major components in the explicit operator for different hardware configurations. MatMul represents the combined wall-time from serial matrix-matrix multiplications by the CPU linear algebra library, and therefore, it is not subject to the OpenMP overhead costs. Flux is composed of large, compute-intensive loop nests that can ameliorate threading overheads with an improved core-to-core performance of Altra relative to POWER9. We find that the relative cost of the Prim component increases going from POWER9 to Altra, but this can be largely attributed to the performance gains in the MatMul and Flux pieces. The differences in the relative costs between the Altra+A100, P9+V100, and TX2+V100 results are all small and likely attributed to differences in the arithmetic intensity for the various kernels.

**Relaxation.** We report the total wall-time to evolve 10 timesteps of the Relaxation benchmark for each hardware configuration in table 3. We measure the improved serial performance of 1.2× (2.2×) for Altra relative to POWER9 (ThunderX2), though it is a smaller improvement than the previous benchmark. The Relaxation benchmark exhibits similar strong scaling efficiency for multi-core performance of Altra, and we find a speedup factor of 1.6 (3.2) relative to POWER9 (ThunderX2). The GPU results are also favorable for the Altra+A100 configuration; we find a 1.7× (1.9×) speedup relative to P9+V100 (TX2+V100) and a 21.5× speedup relative to the Altra CPU-only multi-core case.

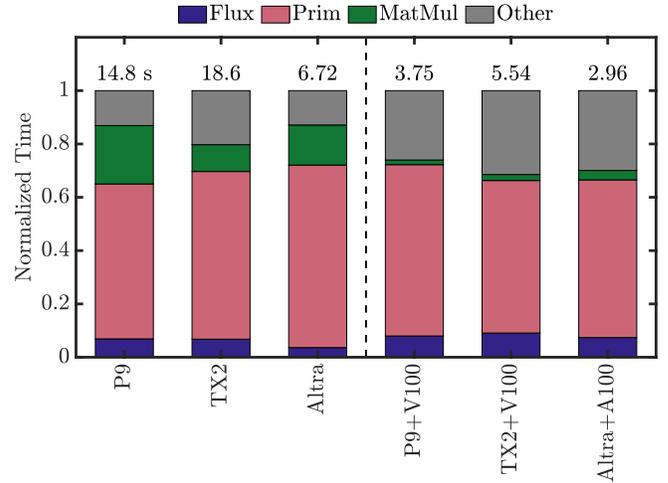

**Fig. 1.** Breakdown of thornado normalized wall-time for components of Streaming Sine Wave test problem. Total absolute wall-clock times are shown above each bar. The CPU-only results are for the configuration of one OpenMP thread per core. All computation is done on the GPU for the V100 and A100 runs; they only use a single CPU thread for managing kernel launches and data movement.

| CPU | GPU | Cores:SMT:Thrds. | Prog. Model | Time (sec) |
|---|---|---:|---|---:|
| Power9 | None | 1:1:1 | OpenMP | 199 |
| ThunderX2 | None | 1:1:1 | OpenMP | 374 |
| Ampere Altra | None | 1:1:1 | OpenMP | 167 |
| Power9 | None | 21:1:21 | OpenMP | 24.6 |
| Power9 | None | 21:2:42 | OpenMP | 25.0 |
| Power9 | None | 21:4:84 | OpenMP | 26.3 |
| ThunderX2 | None | 28:1:28 | OpenMP | 48.9 |
| ThunderX2 | None | 28:2:56 | OpenMP | 46.4 |
| ThunderX2 | None | 28:4:112 | OpenMP | 44.3 |
| Ampere Altra | None | 80:1:80 | OpenMP | 15.3 |
| Power9 | V100 | 1:1:1 | OpenACC | 1.21 |
| ThunderX2 | V100 | 1:1:1 | OpenACC | 1.32 |
| Ampere Altra | A100 | 1:1:1 | OpenACC | 0.71 |

**Table 3.** Comparison of thornado wall-clock times on each platform for the Relaxation test problem. All runs used the nvfortran compiler. Green rows indicate NVIDIA ARM HPC Development Kit hardware.

We can best understand these results by examining the relative cost of the major components in the implicit operator for different hardware configurations in fig. 2. Opacities and Rates are high arithmetic-intensity kernels that can easily realize the improved core-to-core performance of Altra relative to POWER9. In contrast, AA is composed of multiple small loop-nests that incur additional threading overhead. The other components, Update and Prim do not change much between the CPU runs. There is little difference in the relative costs for the Altra+A100, P9+V100, and TX2+V100 configurations.

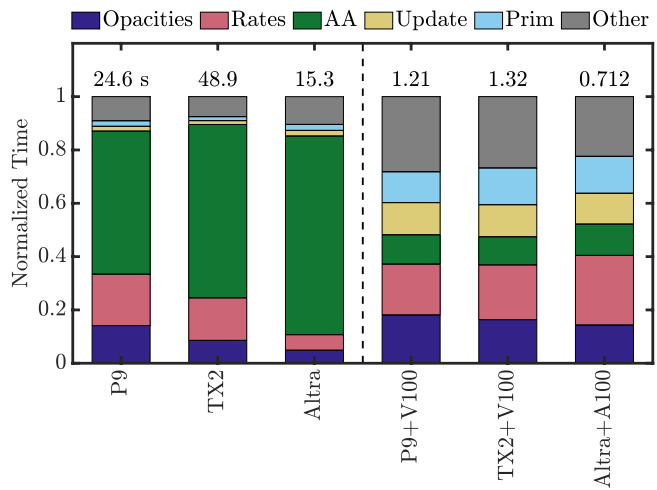

**Fig. 2.** Breakdown of thornado normalized wall-time for components of Relaxation test problem. Total absolute wall-clock times are shown above each bar. The CPU-only results are for the configuration of one OpenMP thread per core. All computation is done on the GPU for the V100 and A100 runs; they only use a single CPU thread for managing kernel launches and data movement.

### 4.2 GPU-I-TASSER

*4.2.1 Background.* GPU-I-TASSER is a GPU-capable bioinformatics method for protein structure and function prediction. It is developed from the Iterative Threading ASSembly Refinement (I-TASSER) method [46]. The I-TASSER suite predicts protein structures through four main steps. These include threading template identification, iterative structure assembly simulation, model selection, and refinement, and the final step being structure-based function annotation. The structure folding and reassembling stage is conducted by replica-exchange Monte Carlo simulations.

I-TASSER has predicted protein structures over the last decade with high accuracy. Thus, it has been ranked as the first automated server for protein structure prediction, according to the critical assessment of structure prediction (CASP) experiments, CASP7 through CASP13 [23].

Despite the robustness of I-TASSER in predicting protein structures with high accuracy, it takes considerably longer to predict some proteins' structures. GPU-I-TASSER has therefore been developed to utilize the efficient GPU in predicting the structure of proteins. GPU-I-TASSER is developed by targeting bottleneck replica-exchange Monte Carlo regions of the protein structure prediction method and porting those to the device. The ported replica-exchange Monte Carlo regions utilize the GPU to optimize the application. The GPU optimization is based on OpenACC parallelization of bottleneck regions with extensive data management.

*4.2.2 Porting for functionality and correctness experience.* GPU-I-TASSER targets the earlier PGI compilers, thus, modifying the makefile, so the compiler and flags are updated to those of nvhpc/22.1, preparing GPU-I-TASSER for use on the testbed. Since GPU-I-TASSER uses OpenACC, utilizing the nvfortran compiler from NVHPC with OpenACC enabled ensures GPU-I-TASSER is ready for testing. Therefore, the build process does not require substantial modifications except updating the makefile with a newer compiler and flags. We made no modifications to GPU-I-TASSER to run on Wombat and observed no issues with correctness.

*4.2.3 Performance and comparisons.* Performance gains across the testbed are compared to the performance from running the same benchmark dataset of proteins on Summit. For details regarding the hardware and software specs of Summit, please refer to [44] To ensure that both systems are on the same level regarding performance comparison, we used the same GPUs. For the initial comparison, we assess the average runtime in seconds for both serial and GPU runs on Wombat using one ThunderX2 processor and one NVIDIA V100 GPU. We observe an average speedup of 7.68× using V100 GPUs on Wombat.

We further compare the performance across V100 GPUs to A100 GPUs on Wombat. We used one A100 and one V100 GPU in this case. We record an average of 7.35× speedup on A100 GPUs compared to the 7.68× on V100 GPUs on Wombat. We should note that the A100 runs were in-comparison to Ampere Computing Altra processors, whereas the V100 performance was relative to ThunderX2 processors. Also, we took the average runtimes against the number of cycles of simulations within a Monte Carlo run.

Finally, we compare the performance of GPU I-TASSER on Wombat to Summit using NVIDIA V100 GPUs. An average speedup of 6.92× is recorded using 1 V100 GPU on Summit. Comparing individual runs on Summit to Wombat, we can observe that Summit performed slightly better than Wombat across GPU and serial runs. Specifically, average serial and GPU runtimes(s) per cycle of simulations are 1669.57 and 217.52, respectively, on Wombat, whereas on Summit, those are 1498.70 and 216.64, respectively.

Figure 3 shows the performance of Wombat's ThunderX2 and Ampere Altra processors and NVIDIA A100 and V100 GPUs relative to the POWER9 processor on Summit. We record a slowdown of an average of 0.9× comparing ITASSER run on Wombat's TX2 processor to Summit's POWER9 processor. For Ampere Ultra, NVIDIA V100, and A100, we record speedup of 1.8×, 6.9×, and 13.3×, respectively.

### 4.3 LAMMPS and Kokkos

*4.3.1 Background.* The Kokkos C++ Programming Model is one of the leading ways of writing performance portable single source code for current and future HPC platforms [43]. It is widely used in the HPC community, particularly within the US National Laboratories and their partners. The programming model is implemented as a C++ abstraction layer on top of vendor-specific programming models such as CUDA, HIP, OpenMP, and SYCL. It is funded by the DOE Exascale Computing Project and developed by a multi-institutional team spanning several DOE laboratories.

LAMMPS is a widely used molecular dynamics application that one can use to simulate a wide range of materials, including condensed matter, gases, and granular materials [42]. It can leverage a wide array of architectures via Kokkos.

*4.3.2 Porting for functionality and correctness experience.* Kokkos already supported various ARM CPUs and all current NVIDIA GPU

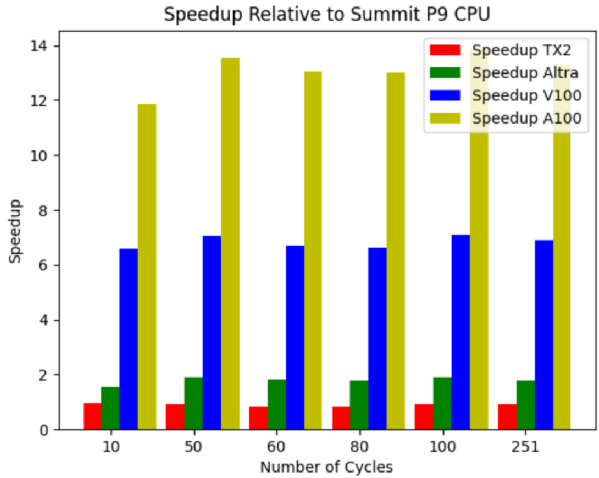

**Fig. 3.** Performance of GPU I-TASSER on Wombat and Summit.

architectures. Thus we required no changes to the build system to build on Wombat. Kokkos-based applications inherit the compiler flags from Kokkos; thus, no changes to the LAMMPS build system were needed. Furthermore, we observed no correctness issues since both Kokkos and LAMMPS already supported ARM and NVIDIA platforms individually.

One significant performance issue for LAMMPS is the current lack of hardware support for GPU-aware MPI on Wombat, which required us to use a LAMMPS runtime option that forces host-only MPI communication. Note that without that restriction, one can achieve significant performance improvements on the Power and X86-based platforms.

*4.3.3 Performance and comparisons.* We decided on four benchmarks that stress host-device interactions to investigate the impact of using ARM as the host CPU for NVIDIA GPUs. Generally, we do not expect code mainly bound by GPU execution time to show different behavior based on the host CPU.

As comparison systems, we used one with an NVIDIA A100 GPU, an AMD EPYC (Milan) X86 CPU, and a system with NVIDIA V100 GPUs and an IBM POWER9 CPU. The latter system connects the GPU and CPU via NVLink. The measured performance numbers are given in table 4.

**Kokkos Kernel Latency.** The Kokkos Programming model provides many different parallel operations, such as `parallel_for` and `parallel_reduce`, which come with different latencies.

Overall, the Wombat system has latencies that fall between the X86 and the IBM POWER-based systems. While the pure launch latencies are comparable to x86, subsequent fences take longer. That, in turn, is reflected in higher latencies for reductions.

**System Atomic Throughput.** To measure the throughput of system atomics, we ran a benchmark distributed as part of the Kokkos repository, which emulates three common atomic access patterns. However, we modified the benchmark to perform the updates into host pinned memory, emulating scenarios where the host and the GPU work on some data collaboratively. The Wombat system performs similarly to the X86 system. The IBM system with NVLink interconnect is significantly faster.

**Host-Device Data Transfer.** We investigate three common host-device data transfer scenarios: transferring data to the device from regular and pinned host allocations and relying on page faults with managed memory.

We fill the host allocation for each scenario and perform a Kokkos `deep_copy` to the device. Subsequently, each value on the device is incremented before copying the data back to the host and then re-setting the values of the host allocation. With managed allocations, `Kokkos::deep_copy` is a no-op, and the time to modify values on the device and host includes the page fault and transfer times, respectively. We use the effective bandwidth of those parallel operations as the data transfer rate for managed allocations.

For regular allocations, all systems perform similarly. With host pinned allocations, Wombat performs 3.5x worse than the IBM system with NVLink, and 25% worse than the X86 system. For managed allocations, the transfer rates depend significantly on the copy direction. Wombat beats the other systems for host-to-device transfers while being the slowest for device-to-host transfers.

**LAMMPS.** LAMMPS demonstrates the impact the observed behavior in the previous micro-benchmarks has on real applications. Often users run small problem sizes per GPU to achieve high simulation rates, making the code kernel latency sensitive. Furthermore, LAMMPS will be impacted by host device data transfer rates due to necessary MPI halo exchanges.

We chose a simple Lennard Jones type simulation with two different problem sizes (32k atoms and 256k atoms per GPU) to demonstrate this sensitivity. We only ran with one and two MPI ranks to avoid conflating the scaling behavior of LAMMPS into the data.

As the micro-benchmark would suggest, the most latency-sensitive scenario (single rank, 32k atoms) performs worse on Wombat than on the X86 system. The larger—less latency sensitive—system performs similarly on Wombat and the X86 system while being slower on the IBM machine due to its older GPU.

When running with two ranks, the total number of kernels increases, resulting in more latency overhead and significant host-device transfers. The data shows that Wombat performs fairly similarly to the X86 system. The IBM system does not seem to benefit from its NVLink connection, indicating that LAMMPS likely uses regular allocations in its non-GPU-aware MPI code path.

## 4.4 MFC

*4.4.1 Background.* MFC (Multi-component Flow Code) is an open-source fluid flow solver available at https://mflowcode.github.io [4]. It provides high-order accurate solutions to a wide variety of physical problems, including multi-phase compressible flows [31] and sub-grid dispersions [3]. MFC employs a finite volume shock and interface capturing scheme via weighted essentially non-oscillatory (WENO) reconstruction, HLL-type approximate Riemann solvers, and total variation diminishing time steppers.Quadrature moment methods handle the sub-grid closures [7].

The MFC codebase is written in Fortran with MPI (and CUDA-aware MPI) capabilities for distributed parallelism. OpenACC provides GPU offloading capability for all compute kernels A Python

Application Experiences on a GPU-Accelerated Arm-based HPC Testbed

| Benchmark | Arm+A100 | x86+A100 | P9+V100 |
|---|---|---|---|
| latency par_for ($\mu$s) | 2.1 | 2.3 | 6.3 |
| latency par_for+fence ($\mu$s) | 10.0 | 8.7 | 15.0 |
| latency par_red ($\mu$s) | 2.3 | 2.7 | 6.2 |
| latency par_red+fence ($\mu$s) | 16.0 | 13.0 | 19.0 |
| atomic histogram (GUp/s) | 0.030 | 0.038 | 0.048 |
| atomic force update (GUp/s) | 0.150 | 0.170 | 0.470 |
| atomic mat.-assembly (GUp/s) | 0.150 | 0.170 | 0.470 |
| transfer h-d regular (GB/s) | 12 | 11 | 12 |
| transfer d-h regular (GB/s) | 11 | 11 | 11 |
| transfer h-d pinned (GB/s) | 18 | 25 | 62 |
| transfer d-h pinned (GB/s) | 15 | 21 | 60 |
| transfer h-d managed (GB/s) | 17 | 11 | 8 |
| transfer d-h managed (GB/s) | 12 | 17 | 26 |
| LAMMPS 1-MPI 32k (MAS/s) | 122 | 148 | 125 |
| LAMMPS 2-MPI 32k (MAS/s) | 95 | 89 | 98 |
| LAMMPS 1-MPI 256k (MAS/s) | 420 | 404 | 320 |
| LAMMPS 2-MPI 256k (MAS/s) | 201 | 201 | 139 |

**Table 4.** Performance of Kokkos-based benchmarks on different platforms. Latencies are measured in microseconds (us), atomic throughput in billion updates per second (GUp/s), transfer rates in GB/s, and LAMMPS performance in million atomsteps per second (MAS/s). Except for latencies, higher is better.

| | # Cores | Compiler | Time [s] | Slowdown |
|---|---|---|---|---|
| NVIDIA A100 | — | NVHPC | 0.28 | Ref. |
| NVIDIA V100 | — | NVHPC | 0.50 | 1.7 |
| 2×Xeon 6248 | 40 | NVHPC | 2.7 | 9.6 |
| 2×Xeon 6248 | 40 | GCC | 2.1 | 7.5 |
| Ampere Altra | 40 | NVHPC | 3.9 | 14 |
| Ampere Altra | 40 | GCC | 2.7 | 9.6 |
| 2×POWER9 | 42 | NVHPC | 4.4 | 16 |
| 2×POWER9 | 42 | GCC | 3.5 | 12 |
| 2×ThunderX2 | 64 | NVHPC | 21 | 75 |
| 2×ThunderX2 | 64 | GCC | 5.4 | 19 |
| A64FX | 48 | NVHPC | 4.3 | 15 |
| A64FX | 48 | GCC | 13 | 46 |

**Table 5.** Comparison of wall-clock times per time step on various architectures. All comparison use either the NVHPC v22.1 or GCC v11.1 compilers as indicated. Highlighted rows indicate NVIDIA ARM HPC Development Kit hardware.

front-end handles input data, execution, and metaprogramming for compiler optimizations. The FFTW package provides access to fast Fourier transforms for computing derivatives in cylindrical coordinates. HDF5 and Silo handle I/O and post-processing.

*4.4.2 Porting for functionality, and correctness experience.* MFC's GPU capabilities require NVHPC compiler version 21.7 or newer and optional support for CUDA-aware MPI. No porting issues were experienced, with support out-of-the-box for the NVIDIA ARM HPC Development Kit hardware using the NVIDIA HPC SDK. We note that better NVIDIA GPU performance is expected with NVHPC compilers, whereas ARM processors like the Ampere Q80-30 shipping with the NVIDIA Development Kits are expected to perform better with ARM or GCC compilers. One must contend with this performance mismatch for workloads that stress GPU and CPU capabilities. However, MFC primarily uses the CPU as a communication and memory manager, so we did not confront this problem.

*4.4.3 Performance and comparisons.* We next investigate the performance of MFC on NVIDIA ARM HPC Development Kits, stressing both the Ampere CPUs and the NVIDIA A100 GPUs. A three-dimensional, two-phase, 16 million grid point fluid dynamics problem served this purpose, representing a typical multiphase flow workload. The performance metric of interest is the average execution wall-clock time over 10 time steps (excluding the first five steps).

We tested performance on several CPUs: Ampere Altra Q80-30 (located on OLCF Wombat), Fujitsu A64FX (OLCF Wombat), Cavium ThunderX2 (OLCF Wombat), Intel Xeon Gold Cascade Lake (SKU 6248, PSC Bridges2), and IBM POWER9 (OLCF Summit). Both NVHPC and GCC11 compilers were tested with -fast and -Ofast compiler optimization flags, respectively. GPU performance was analyzed for the NVIDIA V100 (OLCF Summit) and A100 (OLCF Wombat) using the NVHPC 22.1 compiler with the -Ofast flag. All computations are double precision.

Table 5 shows average wall-clock times and relative performance metrics for the different hardware. The "Time" column has little absolute meaning, with the relative performance being the most meaningful (also shown last column). In table 5 the CPU wall-clock times are normalized by the number of CPU cores per chip. The results show that the A100 GPU is 1.72-times faster than the V100 on OLCF Summit, faster than even the peak double-precision performance would anticipate between the two cards (a factor of 1.24).

A single A100 also gives a 7.3-times speed-up over the fastest tested Intel Xeon Cascade Lake. The GCC11 compiler gives shorter wall-clock times than the NVHPC compiler on all CPU architectures. The Ampere Altra CPUs are 1.4-times faster when compared to the POWER9s and 1.2-times slower than the Intel Xeons. In addition, the ThunderX2 CPUs are about 2-times slower than the POWER9 CPUs. The wall-clock times using the Fujitsu A64FX CPUs are a factor of 10 slower. However, MFC is not explicitly vectorized for ARM instructions. We expect that this and an appropriate Fujitsu ARM compiler are required to extract peak performance from this chip.

Figure 4 shows a time-step normalized breakdown of the duration of the most expensive MFC routines. The left three columns indicate kernel times on GPUs and the rest are CPU-only. When using GPU offloading, all compute kernels are executed by the GPU, with CPU executing I/O and managing halo exchanges. It shows that MPI communications consume a meaningful proportion of the total time on the GPUs but are negligible on CPUs. This result is an artifact of faster routines on the GPUs but approximately constant MPI communication times on CPUs and GPUs. Otherwise, we see that the routine proportions associated with the different CPU and GPU architectures are similar.

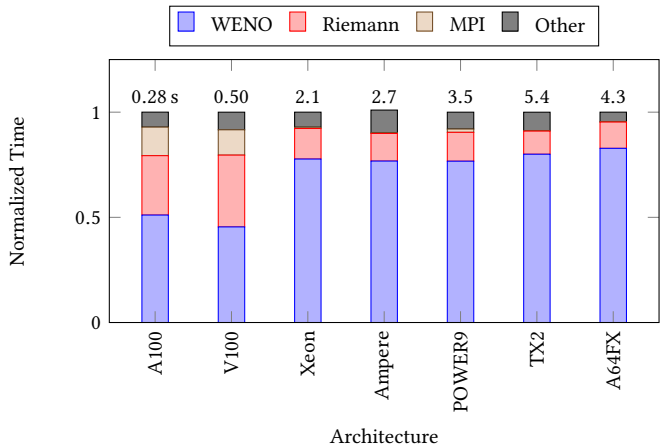

**Fig. 4.** Cost breakdown of different MFC subroutines on various architectures. Cases V100 and A100 have all compute kernels on the respective GPUs, so the associated CPU architecture is not meaningful. Numbers above the bars indicate the absolute wall-clock time (in seconds) as shown in table 5.

### 4.5 MILC

#### 4.5.1 Background.
MILC[10] is an application package concerned with the simulation of Lattice Quantum Chromodynamics (LQCD) to further the study of the (sub-)nuclear physics. MILC handles the generation of gauge field configurations (sampling of the partition function) using Markov Chain Monte Carlo methods, most commonly RHMC [8], and analyzes those configurations to generate physics observables. For both, the dominant algorithm is the iterative linear solver, stemming from the discretized Dirac equation on a 4-d spacetime, giving rise to a sparse matrix, or *stencil*, one must repeatedly solve. Conjugate Gradient is the solver of choice for the commonly used HISQ discretization [12] employed by MILC practitioners.

While popular in the LQCD community, MILC is also often used as a benchmark for HPC systems. Node-level performance is usually dictated by memory bandwidth or, in the case of multi-node scaling, the network bandwidth. Specifically, the inter-process bandwidth must be fast enough to overlay the stencil halo communication with the local stencil application.

MILC runs on GPUs through offload to the QUDA library[11]. For the iterative solver, QUDA utilizes mixed-precision methods to minimize traffic, with a typical linear solver utilizing all double, single, and half precision. Given the propensity for high memory bandwidth on GPUs relative to CPUs, offloading the iterative solver to the GPU dramatically increases the inter-process (GPU) memory bandwidth required to successfully strong scale.

#### 4.5.2 Porting for functionality and correctness experience.
MILC is written in C, employing both OpenMP and MPI parallelism. QUDA is written in CUDA C++: no modifications were required in porting either MILC (commit `d9cc1c9`) or QUDA (commit `4bf4c58`) to the testbed system, and both compiled without issue for the ARM platform using the provided compilers and software stacks (GCC 10.2.0, OpenMPI 4.0.5 and CUDA 11.5.119).

[10]https://github.com/milc-qcd/milc_qcd
[11]https://github.com/lattice/quda

|  | A100 |  |  |  | V100 |  |
|---|---|---|---|---|---|---|
|  | Wombat |  | Rome |  | Summit | TX2 |
| GPUs | 1 | 2 | 1 | 2 | 2 | 2 |
| host | 281 | 170 | 301 | 231 | 462 | 271 |
| compute | 1834 | 1207 | 1878 | 996 | 2133 | 1729 |
| h-d | 75.4 | 39.8 | 68.8 | 46.3 | 76 | 231 |
| d-h | 93.8 | 44.1 | 98.1 | 72.7 | 89 | 63 |
| comms | 163 | 110 | 164 | 99.3 | 213 | 155 |
| other | 203 | 113 | 195 | 103 | 206 | 229 |
| total | 2650 | 1684 | 2705 | 1548 | 3186 | 2645 |

**Table 6.** NERSC MILC Medium Benchmark Time Breakdown (seconds)

Both the Ampere and TX2 nodes on Wombat are limited with respect to inter-GPU communication: their respective host processors do not support the PCIe peer-to-peer protocol, and both systems lack any NVLink GPU interconnect. Thus all communication between GPUs, whether within the same node or between nodes, must be staged through buffers in CPU memory. This limitation also prevents using NVSHMEM, supported by QUDA, so MPI is used exclusively for inter-GPU communication.

#### 4.5.3 Performance and comparisons.
To probe performance, we utilize the NERSC Medium benchmark[12] and look at performance on one and two GPUs on the same node, comparing performance to a platform with AMD EPYC 7742 Rome CPUs and identical A100 GPUs. This platform is similar because it lacks the NVLink interconnect and has the same PCIe gen4 capability. However, critically it supports the peer-to-peer PCIe protocol allowing for inter-GPU communication without staging in CPU memory.[13] We also include measurements taken on the TX2 system compared to Summit, with the latter notably supporting peer-to-peer communication using NVLink. Due to memory footprint size, we include only 2 GPU results.

Table 6 breakdowns the benchmark run times. We note the following key results:

- Single GPU performance is roughly equivalent between Wombat and Rome (2650 s vs. 2705 s), with a slight advantage over Wombat.
- For Dual GPU performance, we see Rome does significantly better (1684s vs. 1548s), with the primary deficit arising due to the "compute".
- The non-GPU accelerated computation "host" shows that Wombat is more than competitive with Rome.
- The raw copy bandwidth between host and device seems to favor the Altra, regardless of the direction of the copy.
- Summit performs significantly better overall than TX2 (2645 s versus 3186 s), with the primary deficit being due to compute.

To better understand the poor scaling of Wombat on two GPUs, in fig. 5 we plot the performance of the HISQ stencil for the three precisions utilized by the mixed-precision solver. The application of this stencil constitutes the bulk of the computing. Without communication, we see performance parity between the two platforms. However, when we include communication overhead, we see that Wombat's performance is severely impacted. In particular, we note that half-precision on 2 GPUs is 45% slower on Wombat versus Rome. We do not include the TX2 and Summit results here for brevity, but we note that a similar picture is painted: with TX2 having a 54% performance deficit for the half-precision stencil.

[12]https://github.com/lattice/quda/wiki/Running-the-NERSC-MILC-Benchmarks
[13]While NVSHMEM is supported on Rome, we chose to make a more direct comparison by deploying MPI exclusively as the communication protocol.

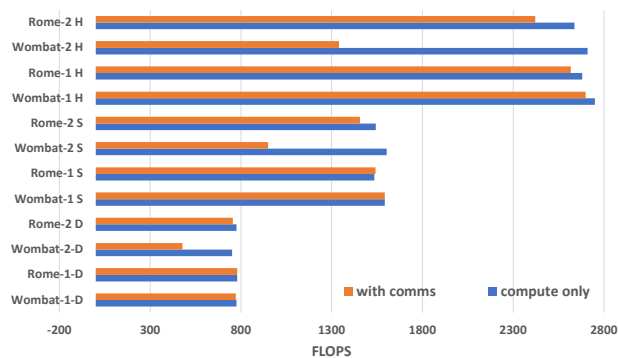

**Fig. 5.** Performance of the QUDA–HISQ stencil with and without overlapping communication. Wombat-1 and Rome-2 denotes Wombat and Rome systems with one A100 GPU. Wombat-2 and Rome-2 denotes Wombat and Rome systems with two A100 GPUs with half (H), single (S), and double (D) precision.

### 4.6 NAMD and VMD

*4.6.1 Background.* NAMD [28] and VMD [18] are biomolecular modeling applications for molecular dynamics simulation (NAMD[14]) and for preparation, analysis, and visualization (VMD[15]). Researchers use NAMD and VMD to study biomolecular systems ranging from individual proteins, large multi-protein complexes, photosynthetic organelles, and entire viruses. Both programs support hardware platforms ranging from personal laptops, workstations, and clouds, up to the largest parallel supercomputers [1]. Researchers perform increasing amounts of VMD simulation preparation, analysis, and visualization work on the same HPC systems where NAMD simulations are run, exploiting high-performance storage systems, parallel analysis, and visualization at scale [15], and avoiding large data transfers [16]. NAMD and VMD are written in C++, C, CUDA, and some platform-specific SIMD vector intrinsics and assembly language for specific performance-critical routines. NAMD is based on the Charm++ parallel runtime system [19], which provides an adaptive, asynchronous, distributed, message-driven, task-based parallel programming model using C++. NAMD and VMD incorporate built-in interpreters for Tcl and Python to provide easy-to-use scripting.

---
[14]https://www.ks.uiuc.edu/Research/namd/
[15]https://www.ks.uiuc.edu/Research/vmd/

*4.6.2 Porting for functionality and correctness experience.* The first adaptations of NAMD and VMD to Arm hardware were performed with SoC on-chip GPU embedded system platforms (NVIDIA CArmA, KAYLA, Jetson TK1, and Jetson TX1), or PCIe-attached GPU (Applied Micro X-Gene/ThunderX + Tesla K20c) system [36]. Wombat presented no compilation barriers for NAMD or VMD, but some minor issues are noted. The Charm++ parallel runtime system used by NAMD did not compile cleanly with GCC 11.1.0, so GCC 10.2 was used to compile NAMD and its associated components. Besides the CUDA toolkit, NAMD also requires FFTW and Tcl libraries, which were easily built on Wombat. Performance results for GPU-resident NAMD are reported in table 7 and table 8.

VMD used a new startup query of CPU SIMD vector instruction set extensions for runtime dispatch of performance-critical loops to hand-vectorized CPU kernels. VMD was extended to query Arm64 CPU vector instruction availability using the Linux kernel `getauxval()` API, enabling runtime detection and kernel dispatch for Arm64 NEON and SVE vector instructions. New hand-vectorized data-parallel NEON and SVE kernels were developed for key atom selection operations and for molecular orbital analysis and visualization, with performance reported in table 9. The new NEON and SVE molecular orbital kernels are direct mathematical and algorithmic descendants from previous CPU and GPU kernels [26, 36–39, 41].

Testing of SVE vector instructions on Fujitsu A64fx nodes demonstrated that two recent versions of the Arm compiler toolchain (21.1 and 22.0) and LLVM (Clang) 10.0.1 generated incorrect code for particular SVE vector intrinsics used in the VMD molecular orbital kernel. As such, the older Arm HPC toolkit version 20.3 was used for the reported results. Similarly, LLVM/Clang versions older than 11.0.1 did not generate correct results for SVE, so the newer version was used for reported results. VMD uses CUDA extensively for performance-critical analysis and visualization tasks. VMD CUDA results are reported in table 10, table 11, and table 12.

VMD supports interactive and offline visualizations on distributed memory HPC platforms, scalable rendering of simulation trajectory movies, and a variety of in-situ visualization tasks. VMD uses OpenGL and EGL for rasterization-based rendering [40], and it uses Tachyon[16] [34], OSPRay[17] [45], and VisRTX[18] (via ANARI [35]) for state-of-the-art ray and path tracing rendering methods on both CPUs and GPUs. Tachyon (CPU and GPU) used the CUDA and OptiX [27] SDKs, and compiled cleanly. VisRTX used the ANARI SDK, CUDA, and OptiX and also compiled without trouble.

*4.6.3 NAMD performance and comparisons.* Benchmarks are shown for the new GPU-resident code path in NAMD [28], which is able to fully utilize an A100 GPU. Although GPU-resident NAMD scales across multiple GPUs on a single node, it depends on high-performance peer-to-peer GPU communication through NVLink using relatively fine-grained load-store operations within CUDA kernels. The lack of this capability on ORNL Wombat limited this study to single GPU performance and the best use of the Ampere Altra.

Two systems are benchmarked representing the extremes of system sizes that are well suited to single-GPU simulation, ApoA1

---
[16]http://www.photonlimited.com/~johns/tachyon/
[17]https://www.ospray.org/
[18]https://github.com/NVIDIA/VisRTX

| CPU :Cores:SMT:Threads | GPU | Comp. | (ns/day) |
|---|---|---|---|
| ThunderX2 : 32:4:2 | V100-PCIe | GCC | 124.9 |
| 2×Power9 : 42:4:7 | V100-NVLINK | XLC | 125.7 |
| 2×Xeon 6134 : 16:2:4 | A100-PCIe | ICC | 181.4 |
| Ampere Altra : 80:1:4 | A100-PCIe | GCC | 182.2 |
| DGX-A100 : 128:2:2 | A100-SXM4 | GCC | 187.5 |

**Table 7.** NAMD single-GPU performance for 92K-atom ApoA1 simulation, NVE ensemble with 12Å cutoff, rigid bond constraints, multiple time stepping with 2fs fast time step, and 4fs for PME. Green rows indicate development kit hardware.

| CPU : Cores:SMT:Threads | GPU | Comp. | (ns/day) |
|---|---|---|---|
| ThunderX2 : 32:4:8 | V100-PCIe | GCC | 9.43 |
| 2×Power9 : 42:4:7 | V100-NVLINK | XLC | 10.26 |
| 2×Xeon 6134 : 16:2:8 | A100-PCIe | ICC | 14.52 |
| Ampere Altra : 80:1:40 | A100-PCIe | GCC | 15.09 |
| DGX-A100 : 128:2:8 | A100-SXM4 | GCC | 15.87 |

**Table 8.** NAMD single-GPU performance for 1M-atom STMV simulation, NVE ensemble with 12Å cutoff, rigid bond constraints, multiple time stepping with 2fs fast time step, and 4fs for PME. Green rows indicate development kit hardware.

| CPU : Cores:SMT:Threads | Compiler | SIMD | Time [s] |
|---|---|---|---|
| 2×Power9 : 42:4:168 | XLC | C++ | 10.60 |
| 2×Power9 : 42:4:168 | XLC | VSX | 6.43 |
| A64fx : 48:1:48 | ArmClang | C++ | 28.88 |
| A64fx : 48:1:48 | ArmClang | NEON | 13.89 |
| A64fx : 48:1:48 | ArmClang | SVE | 4.15 |
| 2×ThunderX2 : 64:4:256 | ArmClang | C++ | 11.12 |
| 2×ThunderX2 : 64:4:256 | ArmClang | NEON | 3.02 |
| Ampere Altra : 80:1:80 | ArmClang | C++ | 5.64 |
| Ampere Altra : 80:1:80 | ArmClang | NEON | 1.35 |
| AMD TR 3975WX : 32:2:64 | ICC | C++ | 19.34 |
| AMD TR 3975WX : 32:2:64 | ICC | SSE2 | 2.89 |
| AMD TR 3975WX : 32:2:64 | ICC | AVX2 | 1.32 |

**Table 9.** Comparison of VMD molecular orbital wall-clock times on each platform. The SIMD column indicates whether compiler autovectorization ("C++"), or hand-coded SIMD vector instructions were used for each test case. Green rows indicate development kit hardware.

| CPU | GPUs | Time [s] |
|---|---|---|
| Cavium ThunderX2 | 1×Tesla V100 | 0.565 |
| Cavium ThunderX2 | 2×Tesla V100 | 0.284 |
| Power9-NVLink | 1×Tesla V100 | 0.394 |
| Power9-NVLink | 2×Tesla V100 | 0.207 |
| Power9-NVLink | 3×Tesla V100 | 0.151 |
| Ampere Altra | 1×A100 | 0.237 |
| Ampere Altra | 2×A100 | 0.126 |
| AMD TR 3975WX | 1×RTX A6000 | 0.234 |
| AMD TR 3975WX | 2×RTX A6000 | 0.125 |

**Table 10.** Comparison of VMD molecular orbital runtime on each platform. Green rows indicate development kit hardware.

(92K atoms) and STMV (1M atoms), and performance is compared with two x86-based configurations, A100–PCIe with Intel Xeon 6134 and A100–SXM4 with AMD EPYC Milan 7763 (a single A100 on DGX–A100). The results are shown in table 7 and table 8, where performance is reported as the number of simulated nanoseconds attainable per day. Each hardware configuration shows the fixed CPU cores and SMT setting together with the number of threads used by NAMD, in which the best performance is achieved when running one thread per core. As the simulated atoms move, the updating of the domain decomposition and rebuilding of device-side data structures are still done on the CPU. The optimal number of threads depends on the size of the system, since adding threads can improve performance up until the thread management overhead exceeds the available computational gain.

| CPU | GPUs | Time [s] |
|---|---|---|
| Cavium ThunderX2 | 1×Tesla V100 | 0.118 |
| Power9-NVLink | 1×Tesla V100 | 0.115 |
| Xeon E5-2660v3 | 1×Tesla V100 | 0.095 |
| Ampere Altra | 1×A100 | 0.061 |

**Table 11.** Comparison of VMD MDFF cryo-EM density map segmentation wall-clock times on each platform. Green rows indicate development kit hardware.

| CPU | GPUs | Time [s] |
|---|---|---|
| Cavium ThunderX2 | 1×Tesla V100 | 0.061 |
| Xeon E5-2697Av4 | 1×Tesla V100 | 0.050 |
| Power9-NVLink | 1×Tesla V100 | 0.049 |
| Ampere Altra | 1×A100 | 0.045 |

**Table 12.** Comparison of VMD MDFF cryo-EM density map quality-of-fit cross correlation wall-clock times on each platform. Green rows indicate development kit hardware.

| CPU | GPUs | DMA+AS (ms) | RT (ms) | Total (sec) |
|---|---|---|---|---|
| Cavium ThunderX2 | 1×V100-PCIe | 87 | 3152 | 3.25 |
| Xeon 6134 | 1×A100-PCIe | 56 | 1920 | 1.98 |
| Ampere Altra | 1×A100-PCIe | 60 | 1671 | 1.73 |
| AMD TR 3975WX | 1×RTX A6000 | 35 | 693 | 0.73 |

**Table 13.** Tachyon GPU ray tracing wall-clock times for each platform. Green row indicates development kit hardware.

The A100–SXM4 configuration proves to be the fastest due to a faster-clocked GPU and PCIe 4.0 bus. The Ampere Altra A100 configuration is the next fastest due to also having a PCIe 4.0 bus. Even though the Ampere Altra cores are SMT 1 and have independent L1 cache memory, performance was improved, especially for the larger system in table 8, by staggering the thread mapping to use just the even-numbered cores. Simulations on A100 are as much as 50% faster than on V100. Similar performance is demonstrated for Cavium ThunderX2 and IBM POWER9, with the latter benefiting from its low latency NVLink connection between CPU and GPU.

*4.6.4 VMD performance and comparisons.* VMD performance results are presented for exemplary analytical and visualization tasks: quantum chemistry molecular orbital analysis and visualization (table 9 and table 10), cryo-EM density map segmentation (table 11), and molecular dynamics flexible fitting cryo-EM density map quality-of-fit (table 12), and GPU ray tracing (table 13).

VMD CPU molecular orbital performance results in table 9 highlight performance achievable with hand vectorization (SIMD vector instruction intrinsics) of complex multiply-nested loops that neither compiler auto-vectorization nor directive approaches are able to vectorize as completely. The hand vectorized kernels permit pervasive vectorization over multiple loop nest levels, through if–then branches, multi-way switch blocks, and across function boundaries without losing efficiency. Each test used all CPU cores and, in most cases, maximum SMT depth to ensure full occupancy of CPU arithmetic ALUs and maximum opportunity for latency hiding.

The GPU-accelerated results in table 10, table 11, and table 12, showcase performance gains provided by much higher peak arithmetic throughput and memory bandwidth of GPUs relative to existing CPU platforms. The GPU molecular orbital results highlight GPU performance and host-GPU interconnect bandwidth.

VMD visualization performance hinges on a combination of host-GPU interconnect latency, bandwidth, and specific GPU hardware acceleration features. The A100 and V100 datacenter GPUs in table 13 use software-based ray tracing, whereas the RTX A6000 has hardware acceleration (RT cores) for ray-triangle intersection and acceleration structure traversal. Hardware-accelerated GPU ray tracing (A40 datacenter, A6000 desktop) performs roughly 3× to 8× faster than software-only [32]. Benefits from rendering on A100 GPUs include large memory capacity, NVLink interconnects, and the opportunity for in-situ visualization to optimize use of limited storage bandwidth. The Xeon (PCIe 3.0) results underperform relative to the Ampere Altra (PCIe 4.0) using the same A100–PCIe GPU. For high resolution (4096×4096) ambient occlusion renderings of a small scene (1M triangles and 0.5M spheres), ray tracing runtimes dwarf host-GPU DMA and ray tracing acceleration structure (AS) build times. As geometric complexity grows, DMA+AS time grows linearly, but ray tracing time grows only logarithmically. For a 156M triangle scene (100× geometry), the DMA+AS time approaches 1 s growing from 5% to 58% of runtime (AMD Threadripper, RTX A6000). The observed RTX A6000 ray casting rate of 7 billion rays/sec ray casting rate is a high fraction of the roughly 10 billion rays/sec peak hardware RT performance.

## 4.7 PIConGPU

### 4.7.1 Background.
PIConGPU [5] is a C++ application that is a scalable, heterogeneous, and fully relativistic particle-in-cell (PIC) code and provides a modern simulation framework for laser-plasma physics and laser-matter interactions suitable for production-quality runs. The code is used to develop advanced particle accelerators for cancer radiation therapy, high-energy physics, and photon science. PIConGPU utilizes the *alpaka* [24] abstraction layer and the particle-in-cell algorithm for its science case simulations. *alpaka* is an open-source abstraction library written in C++17 that aims to provide performance portability across accelerators through the abstraction of underlying levels of parallelism. It is platform-independent and supports concurrent and cooperative use between the host device and any attached accelerators. Alpaka is implemented on top of various accelerator programming APIs like CUDA, HIP, or OpenMP, and therefore, most of the porting is done in alpaka rather than PIConGPU. A recent work narrates efforts to port alpaka to directives-based offloading methods suitable for Arm and GPU platforms [20]. Due to this abstraction layer, only a few top-level changes are made to support PIConGPU running on NVIDIA A100 GPUs via CUDA.

For this work, we use a configuration of PIConGPU that simulates a Weibel instability in a plasma of electrons and positrons, i.e., where all particle species have equal mass. Three variations with different computational intensity are considered: one with a cubic-spline particle shape using single-precision floating point and two with quadratic-splines using single- and double-precision, respectively.

Structurally, PIConGPU is a stencil code with spatial domain decomposition. To facilitate scaling benchmarks, automatic estimation of suitable buffer sizes for particle exchange was introduced into PIConGPU. Each MPI rank exchanges boundary/guard values and particles passing the boundaries with its spatial neighbors using asynchronous point-to-point communication. The particle-grid operations are spatially local and so fit in this scheme.

### 4.7.2 Porting for functionality and correctness experience.
We used Wombat's modules to satisfy PIConGPU's dependencies of a C++17-capable compiler, CMake, CUDA, HDF5, and MPI. When targeting the Arm-based CPUs we used the Armclang++ HPC compiler (version 21.1); for the NVIDIA A100 GPUs we used NVIDIA's nvcc device compiler (version 11.5 included in the nvhpc/22.1 module) together with the gcc/11.1.0 module. In addition, some of PIConGPU's dependencies and recommended libraries (Boost 1.78.0, libpng, pngwriter, and the openPMD) had to be manually compiled because they were unavailable as (suitable) modules.

Compiling PIConGPU for the NVIDIA A100 GPUs worked out-of-the-box without any changes to the code base or the build system. For the Ampere Altra CPUs, it was necessary to add support for Armclang++ to PIConGPU's build system to account for architecture-specific flags, but we required no other changes. We compiled all PIConGPU executables on one of the Ampere compute nodes and supplied the -mcpu=native flag for best machine code generation.

For the following performance evaluation, we used PIConGPU's aforementioned SPEC benchmark configuration and verified the correctness of the results by comparing them to previous benchmark results we have collected on other systems.

### 4.7.3 Performance and comparisons.
Our main analysis focus was execution on Wombat's Ampere nodes since PIConGPU has not been executed on Altra CPUs. Since PIConGPU is not yet a fully heterogeneous code, we did separate runs for the CPUs and the A100 GPUs. Additionally, we evaluated both single precision and double precision data. For all benchmarks, we used the TSC particle form factor. Variation across multiple grid dimensions would result in more MPI overhead, so we restricted the benchmark variants to the $z$ dimension.

**Hardware setup.** We used the same hardware setup for both weak and strong scaling benchmarks. We used one MPI rank per node for the CPU runs, which utilized 80 OpenMP threads (enforced by setting OMP_NUM_THREADS). From PIConGPU's perspective, this constitutes a single CPU device per node, giving us a maximum of eight CPU devices across all Ampere nodes. For the GPU runs, we used two MPI ranks per node with one rank per A100 GPU. For PIConGPU, this setup comprises two GPU devices per node, giving us 16 GPU devices overall.

**Weak scaling.** For the weak scaling analysis, we used a base problem size of 100 time steps and 256 × 256 × 256 cells per computation device. Then we added another 256 cells to the $z$ dimension for any additional device. Table 14 shows the setup per node in more detail.

| Nodes | Devices | Grid layout | Device Layout |
|---|---|---|---|
| 1 | 1 Altra CPU | W: 256 × 256 × 256<br>S: 256 × 256 × 6912 | 1 × 1 × 1 |
| 2 | 2 Altra CPUs | W: 256 × 256 × 512<br>S: 256 × 256 × 6912 | 1 × 1 × 2 |
| 4 | 4 Altra CPUs | W: 256 × 256 × 1024<br>S: 256 × 256 × 6912 | 1 × 1 × 4 |
| 8 | 8 Altra CPUs | W: 256 × 256 × 2048<br>S: 256 × 256 × 6912 | 1 × 1 × 8 |
| 1 | 2 A100 GPUs | W: 256 × 256 × 512<br>S: 256 × 256 × 1024 | 1 × 1 × 2 |
| 2 | 4 A100 GPUs | W: 256 × 256 × 1024<br>S: 256 × 256 × 1024 | 1 × 1 × 4 |
| 4 | 8 A100 GPUs | W: 256 × 256 × 2048<br>S: 256 × 256 × 1024 | 1 × 1 × 8 |
| 8 | 16 A100 GPUs | W: 256 × 256 × 4096<br>S: 256 × 256 × 1024 | 1 × 1 × 16 |

**Table 14.** PIConGPU's scaling setup for Wombat's Ampere nodes. $W$ denotes the weak scaling configuration, $S$ strong scaling.

The results of the weak scaling benchmarks are shown in table 15. With the efficiency staying above 90% for all cases, it can be demonstrated that PIConGPU scales well across multiple Ampere compute nodes – on a previously unknown HPC system and equally unfamiliar hardware – with minimal porting effort.

However, there are also significant differences between CPU and GPU efficiency. This can be explained by the absolute runtime required for the computation as shown in table 17. The GPUs perform the computations much faster than the CPUs. In turn, the GPU weak scaling efficiency is affected by MPI communication overhead much more than the CPU efficiency, likely due to GPU to host data transfer.

| Nodes | Scaling | Altra SP | Altra DP | A100 SP | A100 DP |
|---|---|---|---|---|---|
| 1 | Weak | 1.000 | 1.000 | 1.000 | 1.000 |
| 2 | Weak | 0.998 | 0.997 | 0.992 | 0.986 |
| 4 | Weak | 0.995 | 0.994 | 0.982 | 0.970 |
| 8 | Weak | 0.992 | 0.989 | 0.930 | 0.911 |

**Table 15.** Weak Scaling Efficiency for PIConGPU (where ideal = 1.000). Problem size per device: 256 × 256 × 256 and 100 timesteps. Particle form factor: TSC. SP: single precision, DP: double precision.

**Strong scaling.** For the strong scaling analysis, we used a base problem size of 100 time steps and 256×256×$z$ cells per computation device. $z$ varies between CPUs and GPUs: For CPUs, it is 6912; for GPUs (with less available memory), it is 1024. Table 14 shows the setup per node in more detail.

| Nodes | Scaling | Altra SP | Altra DP | A100 SP | A100 DP |
|---|---|---|---|---|---|
| 1 | Strong | 1 | 1 | 1 | 1 |
| 2 | Strong | 2.00 | 2.04 | 1.89 | 1.92 |
| 4 | Strong | 3.99 | 4.08 | 3.28 | 3.48 |
| 8 | Strong | 7.94 | 8.09 | 4.73 | 5.20 |

**Table 16.** Strong Scaling Factors for PIConGPU (where ideal = N). Problem size per device: 256 × 256 × 256 and 100 timesteps. Particle form factor: TSC. SP: single precision, DP: double precision.

| Nodes | Altra SP | Altra DP | A100 SP | A100 DP |
|---|---|---|---|---|
| 1 | 173.91 s | 209.18 s | 8.56 s | 14.82 s |
| 2 | 174.24 s | 209.79 s | 8.62 s | 15.03 s |
| 4 | 174.78 s | 210.36 s | 8.72 s | 15.27 s |
| 8 | 175.33 s | 211.50 s | 9.20 s | 16.27 s |

**Table 17.** Total computation times for PIConGPU's weak scaling benchmark. Problem size per device: 256 × 256 × 256 and 100 timesteps. Particle form factor: TSC. SP: single precision, DP: double precision.

Table 16 shows the strong scaling speedup achieved by running PIConGPU across multiple nodes. The results corroborate the weak scaling findings: the CPU runs achieve near-perfect speedups when spread across multiple nodes, while the GPU speedups are noticeably below the ideal. In absolute numbers, the GPUs are again much faster than the CPUs (as shown in table 18), so one needs to account for the strong impact of MPI communications.

| # Nodes | Altra SP | Altra DP | A100 SP | A100 DP |
|---|---|---|---|---|
| 1 | 4624.76 s | 5661.73 s | 16.40 s | 29.01 s |
| 2 | 2311.38 s | 2772.75 s | 8.67 s | 15.14 s |
| 4 | 1158.34 s | 1389.25 s | 5.00 s | 8.34 s |
| 8 | 582.00 s | 699.63 s | 3.46 s | 5.58 s |

**Table 18.** Total computation times for PIConGPU's strong scaling benchmark (100 timesteps). Particle form factor: TSC. SP: single precision, DP: double precision.

### 4.8 QMCPACK

*4.8.1 Background.* QMCPACK[21] is an open-source, high-performance Quantum Monte Carlo (QMC) package that solves the many-body Schrödinger equation using a variety of statistical approaches. The few approximations made in QMC can be systematically tested and reduced, potentially allowing the uncertainties in the predictions to be quantified at a trade-off of the significant computational expense compared to more widely used methods such as density functional theory. Applications include weakly bound molecules, two-dimensional nanomaterials, and solid-state materials such as metals, semiconductors, and insulators.

The core components of the application are written using the C++17 standard and targets high-performance computing hardware, and high on-node performance in particular. Synchronous communications demands are low, and MPI scalability of QMCPACK has been demonstrated on tens of thousands of nodes. Most of the computational time used in applications of QMCPACK is spent using the Diffusion Monte Carlo (DMC) algorithm. The compute kernels of this method are common to the other QMC methods implemented in QMCPACK. Therefore, if DMC works well, the other algorithms can also be expected to run efficiently.

The present study's goal is to evaluate the performance of DMC on NVIDIA A100 GPUs and Arm Ampere CPUs using QMCPACK's standard performance tests. They consist of short DMC calculations of variously sized supercells of bulk nickel oxide, *NiO*. The computational cost of these calculations formally scales cubically with the total electron count, which in turn is determined by the atoms in the supercell and their elemental composition.

The DMC algorithm consists of a time stepping loop within which the configurations of a potentially large number of Markov chains (or "walkers") are advanced. This involves particle operations, similar to classical molecular dynamics, and dense linear algebra. The statistical accuracy of the simulation scales with the square root of the number of walkers and total steps. Therefore, QMC methods need to be able to advance as many walkers as possible as quickly as possible. This requires a combination of high memory bandwidth and high floating-point performance.

*4.8.2 Porting for functionality and correctness experience.* Performance portability has been a priority in QMCPACK, with algorithms continuously being thoroughly tested on x86_64 CPUs and NVIDIA and AMD GPUs using OpenMP, CUDA, and ROCm/HIP programming models and the highly optimized BLAS vendor libraries. QMCPACK relies on the CMake build system to find appropriate dependencies and configurations. It also has an extensive set of unit and integration tests. As such, compilation on Wombat was essentially a straightforward process. We prioritized building with the latest vendor compiler available on Wombat: ArmClang and NVHPC for CPU and GPU runs, respectively. Nevertheless, the ArmClang shipped CPU BLAS implementation using OpenMP did not work, throwing a runtime exception.

*4.8.3 Performance and comparisons.* We set up a set of problem sizes in the *NiO* supercell benchmark characterized by the number of electrons in the system. Memory usage is formally quadratic in the electron count. As memory requirements increase, the number of potential "walkers" that can fit in the GPU or on-node memory reduces. Because the GPU implementation batches work over the number of walkers, the achievable efficiency can be limited if the batch size can not be large enough before the GPU memory is exhausted.

Performance is measured using a throughput metric. As defined in (1), Throughput is a measurement for the computational cost associated with a single DMC simulation yielding to the frequency of advancing walkers in the DMC simulation, with higher values indicating better performance. The cost is formally cubic in the electron count and linear in the walker count. Thus the throughput drops dramatically at large electron counts.

$$\text{Throughput} = \frac{\text{walkers} \times \text{blocks} \times \text{steps}}{\text{DMC time}} \quad (1)$$

**GPU-only Results.** The initial focus on targeting Wombat's NVIDIA's A100 GPUs on Ampere nodes is to understand the number of possible "walker count per GPU device" for the *NiO* supercell benchmark for different system sizes. Walker counts in QMCPACK are equivalent to the "batch size" for GPU computation, finding the maximum number of walkers also allows for efficient use of each available GPU. We apply a bisectional search to find the maximum walker count limits due to memory limitations within a single walker count range for accuracy (+/−1 walkers). The resulting walker count limits per A100 GPU, offering 40 GB of memory, are given in table 19 which also provides this information for reference on the V100 GPU, offering 16 GB of memory, from our experiments on Summit. As the system size increases, the benefits of the A100 memory become larger, with the largest measured system size of 6144 electrons surpassing the simple memory ratio between A100 and V100 of $2.5x = 40/16$ by a factor of 32 due to the significant additional memory overheads in storing wavefunctions used in the calculation.

| NiO supercell electrons | max walkers Summit V100 | max walkers Wombat A100 |
| ---: | ---: | ---: |
| 48 | 65535 | 65535 |
| 96 | 35419 | 65534 |
| 192 | 12554 | 32797 |
| 384 | 818 | 2047 |
| 768 | 785 | 2047 |
| 1152 | 423 | 1244 |
| 1536 | 240 | 719 |
| 2304 | 96 | 322 |
| 3072 | 43 | 174 |
| 6144 | 1 | 32 |

**Table 19.** The maximum number of walkers (batch size) on a single Wombat A100 and Summit V100 GPU.

We use the walker count on table 19 on each system to compare the DMC performance throughput on (1) ranging from 1 GPU to the maximum limit using Summit's 6 V100 GPUs and Wombat's 2 A100 GPUs per node. Results are illustrated in fig. 6 showing the results obtained on Wombat using the NVHPC compiler and on Summit. As expected, single A100 GPU runs on Wombat outperform those on V100s, with significantly larger throughput for nearly all problem sizes. When using all the available GPUs per node on each system, we observe that for smaller cases, Summit 6 V100 GPUs outperform in terms of throughput per node. However, Wombat's A100 2 GPUs are significantly more performant for the largest and most computationally challenging case. For these system sizes, greater GPU memory is the biggest factor in increased performance.

**CPU-only Results.** QMCPACK CPU configuration assigns walkers to individual OpenMP threads. Within each step, they are advanced independently. We obtained initial performance results for a single walker, single thread run on Wombat's Ampere nodes and compared with similar configurations on Intel Xeon 6248R and Summit POWER9 nodes. As shown in fig. 7, the Ampere runs using the ArmClang21–OpenBLAS configuration is highly competitive and are highest in performance for the largest electron count. Nevertheless, while there is potential for the Arm configuration, we observed a significant degradation in OpenMP scalability shortly after the number of threads increases > 4, traced to limitations in the BLAS libraries. The latter serves as an opportunity for the Arm-provided compiler and performance libraries. In addition, aspects such as runtime bugs and performance degradation have been communicated to the compiler teams, and we look forward to seeing benefits from the Arm software stack.

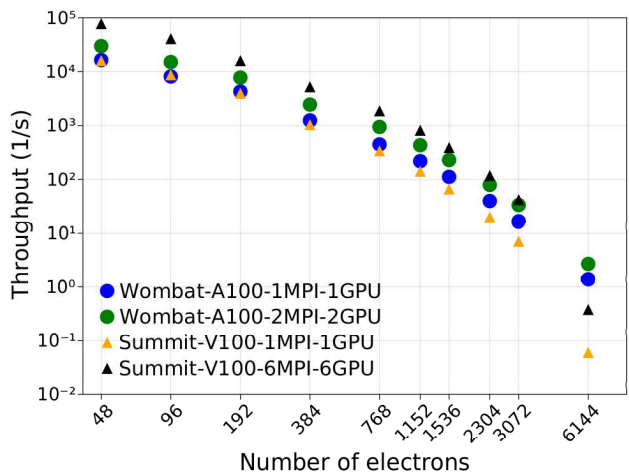

**Fig. 6.** QMCPACK DMC throughput for Wombat and Summit nodes as a function of the number of electrons in the *NiO* benchmark from table 19.

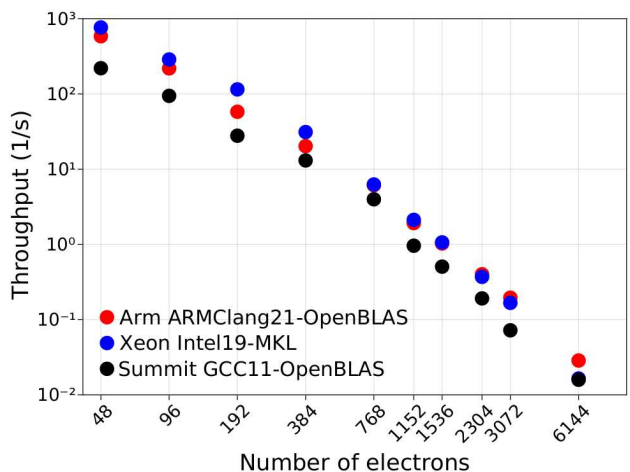

**Fig. 7.** QMCPACK DMC throughput for a single walker/core across different CPU systems (Wombat, Xeon, Summit) as a function of the number of electrons in the *NiO* benchmark.

### 4.9 SPEC HPC 2021

*4.9.1 Background.* SPEChpc 2021 is a benchmark suite comprised of real-world application codes designed for portable performance across heterogeneous CPU and GPU architectures [2][19]. SPEChpc provides C/C++ and Fortran codes, accelerated by OpenMP, OpenMP TGT Offloading, OpenACC, and CUDA programming models. On Wombat, we utilized SPEChpc 2021 to evaluate single-node performance using one to two NVIDIA A100 GPUs while varying the number of cores bound to each GPU.

---
[19] https://www.spec.org/hpc2021/

*4.9.2 Porting for functionality and correctness experience.* As a benchmark designed for portability, SPEChpc required minimal porting effort, aside from identifying the correct compiler flags and launch commands to use. SPEChpc uses a custom harness to build and launch codes, which requires a simple configuration file that specifies compiler flags and launch commands. We targeted the NVHPC and LLVM compilers, as NVPHC provides both OpenACC and OpenMP target offloading, and LLVM provides OpenMP target offloading.

*4.9.3 Performance and comparisons.* We ran the SPEChpc 2021 suite on Wombat (Ampere N1 + NVIDIA A100) nodes, with comparisons to ORNL's Summit (IBM POWER9 + NVIDIA V100) nodes. The compilers used on Wombat were NVHPC 22.1 using OpenMP target offloading (TGT) and OpenACC offloading (ACC), and LLVM v15.0.0 using OpenMP target offloading (TGT). LLVM is not built with Fortran support, so the POT3D, SOMA, and Weather benchmarks are not run with LLVM. Three iterations of the *tiny* benchmark were performed on Wombat. On Wombat, we tested with combinations of one and two NVIDIA A100 GPUs. We ran the benchmark suite using one and two ranks per GPU for a total of four data points for each acceleration model. On Summit, we tested the use of six V100 GPUs with one iteration using one rank per GPU. Summit displays several runtime errors while running on one V100 GPU because the SPEChpc *tiny* benchmark targets about 40GB of memory usage, which exceeds the V100 limit of 16GB, and is why one V100 is not shown in the following results.

Figure 8 and fig. 9 display the performance (measured as walltime) of the OpenMP target offloading implementations of NVHPC and LLVM on Wombat and Summit, respectively, relative to NVHPC OpenACC. A 19x difference in runtime is observed in Minisweep from NVHPC-ACC to NVHPC-TGT on Wombat using a single GPU, one rank per GPU, and a 14x difference is observed when using both A100 GPUs. This behavior is not limited to Wombat, as Summit also observed an 8x slowdown from NVHPC-ACC to NVHPC-TGT when using all 6 GPUs, one rank per GPU. This behavior is also not limited to NVHPC's OpenMP offloading, as LLVM-TGT demonstrates a 4-6x slowdown on Minisweep on both Summit and Wombat.

Using one GPU on Wombat, five of the six codes that complete with NVHPC–TGT are slower than when using NVHPC–ACC, and all three of the codes that complete for LLVM-TGT are slower than when using NVHPC-ACC. On all GPUs, 7 of the 9 codes run faster using ACC than TGT on Wombat, and 5 of the 7 codes that complete without a runtime error on Summit run faster using ACC than TGT.

### 4.10 SPH-EXA2

*4.10.1 Background.* The SPH-EXA2 project is a multidisciplinary effort that extends the SPH-EXA[6] project and that looks to scale the Smoothed Particle Hydrodynamics (SPH) method to enable exascale hydrodynamics simulations for the fields of Cosmology and Astrophysics. On Wombat, we used the Sedov-Taylor blast wave explosion test [13] to simulate a spherical shock generated by the instantaneous injection of thermal energy at a single point in a static uniform background. This test requires the code to simulate shock-fronts while correctly maintaining spherical symmetry and conservation laws.

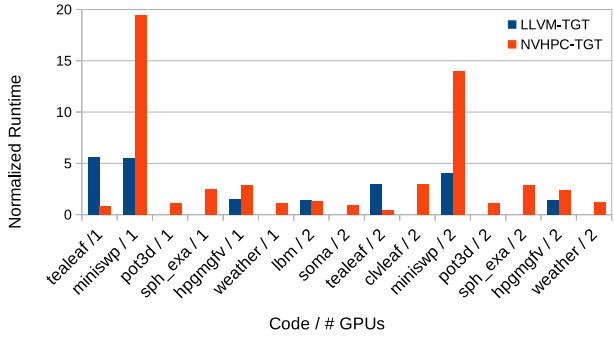

**Fig. 8.** Performance of SPEChpc 2021 on Wombat using OpenMP Target (TGT) offloading, relative to OpenACC.

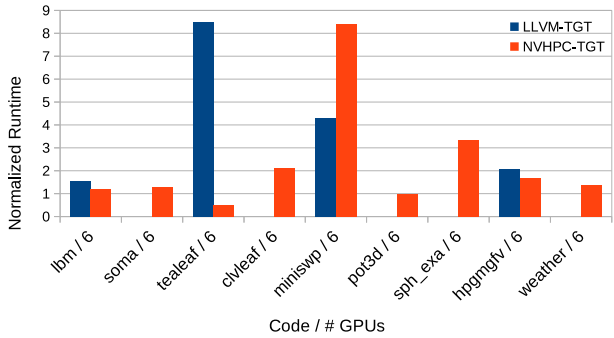

**Fig. 9.** Performance of SPEChpc 2021 on Summit using OpenMP Target Offloading (TGT) offloading, relative to OpenACC.

SPH-EXA2[20] is open source, written in C++17, parallelized with MPI and OpenMP, and accelerated with CUDA or HIP.

*4.10.2 Porting for functionality and correctness experience.* Compiling SPH-EXA2 on Wombat was straightforward with the right CPU (-mcpu) and GPU (-arch) compiler flags including (`neoverse-n1` for Ampere, `thunderx2t99` for TX2, `a64fx` for A64FX and `sm_80` for A100).

We used the ReFrame framework to build, run and analyze the code. For correctness, we measured the convergence of the Sedov-Taylor test. We observed no correctness issue: the measured L1-norm errors for density, pressure and velocity (0.15792, 0.91919, and 0.92395) with the GNU/11 compiler and (0.15924, 0.92013, and 0.92627) with the ARM/21 compiler on the three different AArch64 CPUs (N1, TX2, and A64FX) are comparable with results on x86_64 CPUs.

Additionally, we ran the unit tests of the SPH-EXA2 code-base to ensure there are no problems with the functionality on the Wombat platform, and we report no problems have been encountered.

*4.10.3 Performance and comparisons.* To investigate the impact of using the ARM CPU on SPH-EXA2, we report the performance results on three different systems within the Wombat platform and compare the results with non-ARM systems of a CPU-only run by the results of a CPU+GPU run using a single node. Table 20 describes the CPUs of Wombat and those of two other x86_64 systems used for comparison.

---

[20] https://github.com/unibas-dmi-hpc/SPH-EXA

|   | CPU cores | Threads/ core | NUMA domains |
|---|---|---|---|
| Ampere Altra N1 | 80 | 1 | 1 |
| Marvell TX2 | 64 | 4 | 2 |
| Fujitsu A64FX | 48 | 1 | 4 |
| AMD EPYC 7662 Zen2 | 128 | 2 | 4 |
| Intel CLX 6258R | 112 | 2 | 2 |

**Table 20.** SPH-EX2 CPU Configuration.

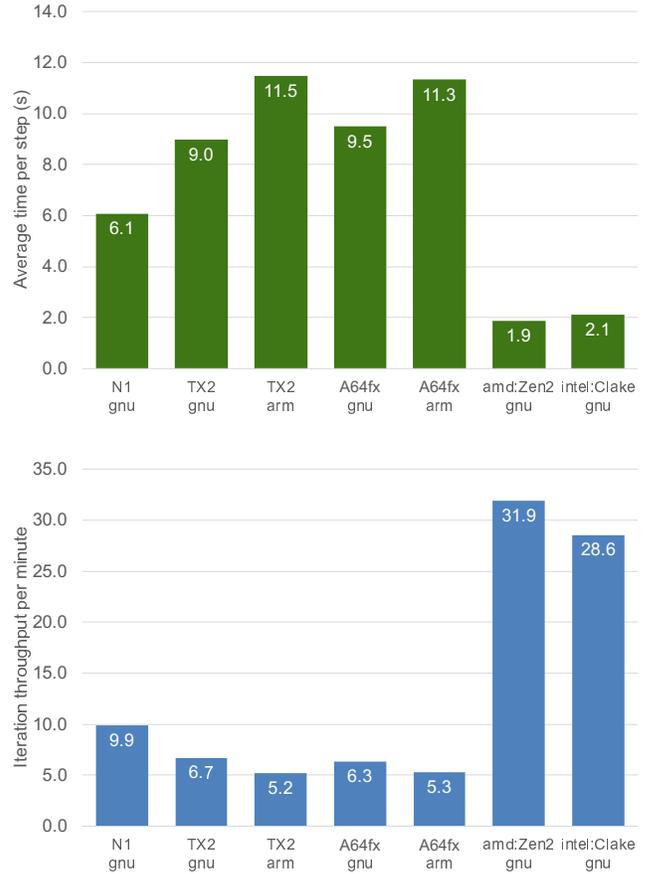

**Fig. 10.** SPH-EXA2 execution using MPI+OpenMP on the CPU-only setup with $200^3$ particles and 800 time-steps for the Sedov-Taylor blast wave explosion test.

**CPU-only Results.** Figure 10 shows the results obtained for SPH-EXA2 code executing the Sedov–Taylor blast test case with $200^3$ particles using MPI+OpenMP on CPU only setup. The average time in seconds per time step of the simulation is shown on the top chart (lower is better), and the achieved iteration throughput per minute of the simulation is shown on the bottom chart (higher is better) of the figure. On Wombat, the best performance is obtained with the GNU compiler on the Ampere N1 CPU, while the overall best performance is achieved on x86_64 CPUs. Systems with fewer cores per socket (Table 20) lead to overall lower performance than those with higher core counts. Additionally, the results on TX2 and A64FX systems show that the SPH-EXA2 code compiled with the GNU compiler performs better than the ARM compiler.

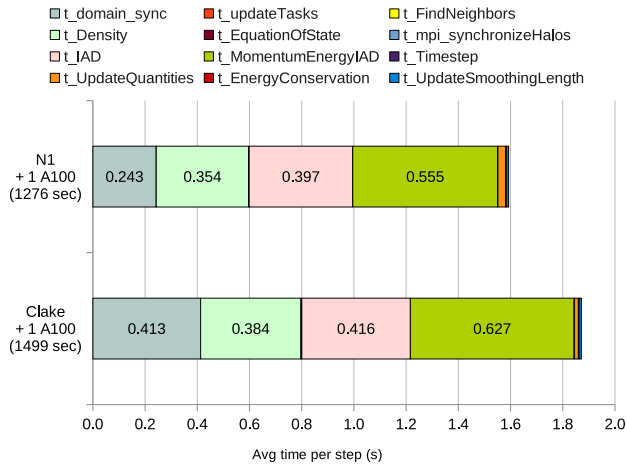

**Fig. 11.** Execution times of SPH-EXA2 executing the Sedov-Taylor blast test (MPI+OpenMP+CUDA, CPU+GPU) for 800 time-steps with $200^3$ particles, using 1 NVIDIA A100-PCIe-40GB per compute node.

| sph-exa sedov-cuda -n200 -s800 | HtoD N1 | HtoD Clake | DtoH N1 | DtoH Clake |
|---|---|---|---|---|
| Size (GB) | 1744 | 1744 | 1488 | 1488 |
| Time (s) | 134 | 302 | 125 | 214 |
| Bandwidth (GB/s) | 13.0 | 5.8 | 11.9 | 7.0 |

**Table 21.** GPU: CUDA memcpy operations between host and device

Further code profiling using the ARM Performance Reports tool allowed us to identify the cause of the difference in performance between N1 and A64FX CPUs since the former has fewer cores but performs better in our tests. Profiling showed that a higher number of L2 cache misses and stalled cycles on the A64FX CPUs cause performance to degrade in these systems. We believe this is due to N1 having only 1 NUMA node compared to the 4 NUMA nodes of A64FX. Further analysis is needed to use the vectorization support (SVE) better and increase compute performance.

**CPU+GPU Results.** Figure 11 shows the execution times of the MPI+OpenMP+CUDA version of the SPH-EXA2 code for the Sedov–Taylor blast test case with $200^3$ particles and 800 time-steps. The N1 system on Wombat slightly outperforms the x86_64 reference system. The difference in performance is caused by the Wombat N1 having PCIe 4.0 compared to the x86_64 reference system's PCIe 3.0 port, which creates the difference between data transfer rates between the CPU and the GPU resulting in the N1 system achieving higher overall execution performance. The size and speed of CUDA memcpy operations reported in table 21 show that the same amount of data was transferred between host (H) and device (D) on both systems, with higher transfer rates on Wombat's N1.

After using Nsight, SPH-EXA2's top kernels can be identified as compute-bound, and the measured performance shows that using ARM as the host CPU has no negative impact on the execution time of the kernels.

## 5 RELATED WORK

Prior work has primarily focused on the evaluation of HPC applications on the Arm Cavium ThunderX2 with the Aries interconnect as part of the Isambard supercomputer [25] and the A64FX processor with TOFU interconnect in the Fugaku system [30] and with Infini-Band interconnect [11] on the Okami system. Other related work has looked at Arm-based performance portability with TX2 and previous generation Ampere nodes [9] and concludes that Kokkos and OpenMP provide performance portability across Arm and x86 platforms. A more recent update adds SYCL evaluation but comes to similar conclusions [10].

In terms of more cloud-HPC-focused efforts, a recent hackathon run by the non-profit Arm HPC User Group, AWS, and Arm supported the testing and development of HPC codes on AWS's custom Graviton2 instances. This event, the AHUG Hackathon: Cloud Hackathon for Arm-based HPC [21], supported 30 teams to investigate the top HPC applications used on AWS and helped test Spack packages with flags for the Graviton2 setup as well as Reframe testing scripts for Arm and x86 platforms. The effort focused on porting several HPC applications running on Arm, including a full set of mini-apps and applications [22], but it did not include any accelerated nodes. This work complements other HPC application efforts on AWS, including Nalu [23], a CFD modeling code, and NWChem [24], a widely used quantum chemistry code.

A recent evaluation effort of the SPEChpc 2021 suite [2] also includes similar application evaluations for platforms including x86 CPUs and NVIDIA and AMD GPUs. This Arm-based investigation contributes complementary results to the SPEChpc study across a mostly different set of applications and benchmarks. All previous Arm application study efforts do not include the evaluation of Arm with GPU systems.

## 6 CONCLUSIONS

In this work, we used the Wombat testbed at the Oak Ridge Leadership Computing Facility (OLCF) to study the readiness and usability of a modern GPU-accelerated Arm-based HPC platform, the NVIDIA Arm HPC Developer Kit. Ten representative applications from different scientific domains, and using a variety of programming models and languages were selected, built on the platform and tested for correctness. Wherever possible, performance was compared with other leading HPC platforms used for production science, as well as other Arm-based platforms that are part of the Wombat system.

As can be seen from the various applications experiences, the porting process was straightforward, and mostly required minor modifications to the build systems to be able to compile and run on the target platform. The availability of a fairly mature set of compilers that cover the gamut of used programming models was crucial in achieving this seamless porting process. Of particular note, the availability of the NVIDIA HPC SDK facilitated the porting process or those applications that currently use this tool-chain on other GPU-accelerated supercomputers such as Summit. Furthermore, the maturity of Arm support in the spack package management system greatly facilitated the deployment of third party tools and libraries needed by the various application teams.

---

[21]https://community.arm.com/arm-community-blogs/b/high-performance-computing-blog/posts/aws-arm-ahug-hpc-cloud-hackathon
[22]https://github.com/arm-hpc-user-group/Cloud-HPC-Hackathon-2021/tree/main/Applications
[23]https://community.arm.com/arm-community-blogs/b/high-performance-computing-blog/posts/low-mach-number-cfd-modeling-with-nalu-on-graviton2-aws-m6g
[24]https://www.youtube.com/watch?v=xq_sj4nAk3k

While exhaustive performance optimization was not a primary goal of this work, we carried out preliminary performance measurements to assess the overall platform readiness. For application considered *GPU-dominant*, performance improvements were commensurate with the hardware capabilities of the NVIDIA Ampere GPU (A100) relative to the previous generation NVIDIA Volta GPU (V100) and using an Arm-based CPU did not adversely impact the outcome. For a subset of the applications where the code can be configured to run only on the CPU, we carried out several CPU-only experiments and we observed that the performance of the Ampere CPU was generally competitive with leading X86-64 and Power9 CPUs. It should be noted that the lack of an appropriate fast and fully RDMA-capable CPU-GPU bus in the Wombat testbed (similar to NVIDIA NVLink on POWER9 CPU in Summit, or AMD's xGMI in the newly installed Frontier supercomputer at OLCF) and the lack of NVLINK across the CPU and GPUs adversely impacted performance for applications that require fast data movement across the different processing elements in the platform. Exploiting these features requires a holistic design that combines needed system software with a hardware design that adopts a GPU-centric platform design. Such a design can be found in systems such as NVIDIA DGX[25] or Frontier[26], where the GPUs are connected directly to the NICs on the node. In the near future more tightly integrated cache-coherent CPU-GPU platforms (e.g. NVIDIA Grace Hopper Superchip) will further enhancing developers productivity and platform programmability.

Evaluating testbeds is a continuous process. We plan, as our next step, to investigate the Arm platform's usability for large data and machine learning workloads as well as the exploitation of NVIDIA BlueField Data Processing units (DPU). As more Arm-based platforms from various vendors become available in the market, we anticipate continuing this evaluation effort to gain a more thorough understanding of the platform's strengths and potential incompatibilities with different classes of applications.

## ACKNOWLEDGMENTS

This research used resources of the Oak Ridge Leadership Computing Facility at the Oak Ridge National Laboratory, which is supported by the Office of Science of the U.S. Department of Energy (Contract No. DE-AC05-00OR22725). Assessment of QMCPACK and ExaStar was supported by the Exascale Computing Project (17-SC-20-SC), a collaborative effort of the U.S. Department of Energy Office of Science and the National Nuclear Security Administration.

VMD and NAMD work is supported by NIH grant P41-GM104601. S. H. Bryngelson acknowledges the use of the Extreme Science and Engineering Discovery Environment (XSEDE) under allocation TG-PHY210084, OLCF Summit allocation CFD154, hardware awards from the NVIDIA Academic Hardware Grants program, and support from the US Office of Naval Research under Grant No. N000142212519 (PM Dr. Julie Young). E. MacCarthy acknowledges Yang Zhang of University of Michigan, Ann Arbor, for providing the I-TASSER code. Work on PIConGPU was partially funded by the Center of Advanced Systems Understanding which is financed by Germany's Federal Ministry of Education and Research and by the Saxon Ministry for Science, Culture and Tourism with tax funds on the basis of the budget approved by the Saxon State Parliament. The work in SPH-EXA2 is supported by the Swiss Platform for Advanced Scientific Computing (PASC) project SPH-EXA2 (2021-2024) and as part of SKACH consortium through funding from the Swiss State Secretariat for Education, Research and Innovation (SERI).

---

[25]https://www.nvidia.com/en-au/data-center/dgx-systems/
[26]https://olcf.ornl.gov/wp-content/uploads/Frontiers-Architecture-Frontier-Training-Series-final.pdf